\documentclass{article}
\usepackage{graphicx}
\begin{document}
\textheight 9in
\topmargin -.5in
\vbadness = 100000
\hbadness = 100000
\title{\mbox{The lack of rotation in a moving right angle lever}}
\author{Jerrold Franklin\footnote{Internet address:
Jerry.F@TEMPLE.EDU}\\
Department of Physics\\
Temple University, Philadelphia, PA 19122-6082}
\maketitle
\begin{abstract}
The absence of any tendency toward rotation in a moving right angle lever
is given a simple explanation.  
\end{abstract}

One of the earliest paradoxes of special relativity is the ``right angle lever paradox", first proposed in 1909 by G. N. Lewis and R. C. Tolman\cite{lt}.  The paradox involves a right angle lever at rest in a system S, acted on by four forces as shown in Fig.\ 1.  
The lever is in static equilibrium in its rest system with
the vector sum of the forces zero, and the sum of moments $\bf r\times F$ zero,  with no tendency to translate or rotate.  Each force $\bf F$ is here interpreted as being equal to the resulting $\frac{d{\bf p}}{dt}$ if that $\bf F$ were the only force on the object.  This is the usual interpretation of the ``Lorentz force".  If a Lorentz transformation is made to a system S$'$ moving with velocity $-{\bf v}$ parallel to $\bf r_\parallel$, it is found that the sum of moments $\bf r'\times F'$ does not vanish.  The implication is then made that this would lead to a rotation of the lever, seemingly violating the postulate of special relativity that a uniform velocity translation should not change the physics.  This result constitutes the Lewis-Tolman right angle lever paradox.
\begin{center}
\includegraphics[width=4in]{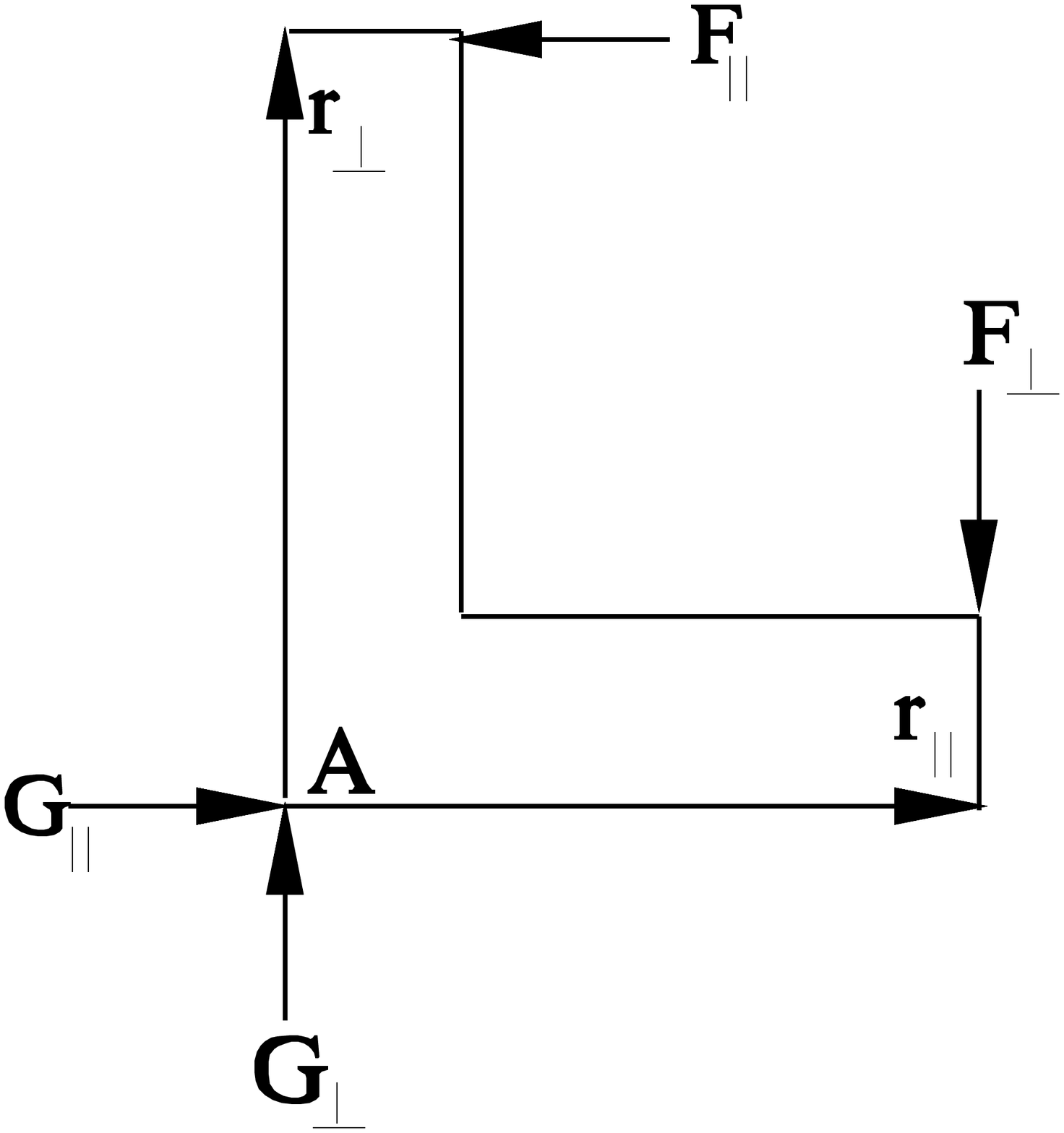}
\end{center}
\noindent
\mbox{{\bf Fig.1:}  Forces acting on a right angle lever in static equilibrium in its rest system.}\\   
\\

The paradox has been treated in the literature[2-4] and has more recently received a great deal of interest on various internet sites.  The published and unpublished discussions have introduced a wide variety of explanations, generally introducing {\it ad hoc} effects of questionable validity to keep the moving lever from rotating.
\footnote{We remark on the published analyses in the Appendix of this paper.}
Because students on all levels can become confused, especially by the internet discussions, it is particularly important that paradoxes such as the moving lever be treated correctly in undergraduate and graduate courses.
  
In this note we point out that the analysis of the similar motion of a capacitor in our recent paper ``The lack of rotation in the Trouton-Noble experiment"\cite{tn} can be applied to show that a moving right angle lever does not rotate, even though the sum of $\bf r'\times F'$ (with $\bf F'$ interpreted as above) does not vanish.  For the Trouton-Noble capacitor, the sum $\bf r'\times F'$ also does not vanish, but there is no rotation because 
$\frac{d{\bf p}}{dt}$ is not parallel to the acceleration that would result.  That is, the relativistic connection between $\frac{d{\bf p}}{dt}$ and the acceleration, defined as
${\bf a}=\frac{d{\bf v}}{dt}$, is (We use units with $c$=1.)
\begin{equation}
\frac{d\bf p}{dt}=\frac{d}{dt}(m{\bf v}\gamma)
= m\frac{d}{dt}\left[\frac{\bf v}
{\sqrt{1-{\bf v}^2}}\right]
=m\gamma^3[{\bf a}+{\bf v\times(v\times a)}].
\end{equation}
Because any ensuing motion would be in the direction of $\bf a$, it is 
$\bf r\times a$, and not ${\bf r}\times\frac{d{\bf p}}{dt}$ that determines whether a force will produce rotation.

We first calculate the sum of $\bf r'\times F'$ for the primed system, where the lever is moving with velocity $\bf v$.
For the case of the lever in Fig.\ 1, we take moments about the point A.  This eliminates any torque due to the forces
$\bf G_\perp$ or $\bf G_\parallel$, which act through point A.
In the lever's rest system,
\begin{equation}
{\bf r_\parallel\times F_\perp}
+{\bf r_\perp\times F_\parallel}=0,
\end{equation}
corresponding to static rotational equilibrium.

The coordinates transform as
\begin{eqnarray}
{\bf r'_\parallel}&=&{\bf r_\parallel}/\gamma\\
{\bf r'_\perp}&=&{\bf r_\perp},
\end{eqnarray}
where ${\bf r_\parallel}$ is parallel to $\bf v$ and ${\bf r_\perp}$ is perpendicular.
To transform the force, we use the four-vector Minkoswki force, defined as
\begin{equation}
{\cal F}^\mu=(\gamma{\bf F\cdot v},\gamma{\bf F}).
\end{equation}  
This leads to
\begin{eqnarray}
{\bf F'_\parallel}&=&{\bf F_\parallel}\\
{\bf F'_\perp}&=&{\bf F_\perp}/\gamma.
\end{eqnarray}
Then, the sum of moments in S$'$ is
\begin{equation}
{\bf r'_\parallel\times F'_\perp}
+{\bf r'_\perp\times F'_\parallel}
={\bf r_\parallel\times F_\perp}/\gamma^2
+{\bf r_\perp\times F_\parallel}
={\bf -v}^2({\bf r_\parallel\times F_\perp}),
\end{equation}
which does not vanish.

We now show that if there is no tendency to rotate
in the lever's rest system, there will be no tendency to rotate in the system S$'$ where the lever is moving with velocity $\bf v$, if ``tendency to rotate" is properly interpreted.
The condition for no tendency to rotate in the lever's rest system is
\begin{equation}
{\bf r_\parallel\times a_\perp}
+{\bf r_\perp\times a_\parallel}={\bf 0}.
\end{equation}
The relativistic transformation equations from the rest system to S$'$ for the accelerations are
\begin{eqnarray}
{\bf a'_\parallel}&=&{\bf a_\parallel}/\gamma^3\\
{\bf a'_\perp}&=&{\bf a_\perp}/\gamma^2.
\end{eqnarray}
The sum of moments in system S$'$ is given by
\begin{eqnarray}
{\bf r'_\parallel\times a'_\perp}
+{\bf r'_\perp\times a'_\parallel}
&=& \left(\frac{\bf r_\parallel}{\gamma}\right)
\times\left(\frac{\bf a_\perp}{\gamma^2}\right)
+\left({\bf r_\perp}\right)
\times\left(\frac{\bf a_\parallel}{\gamma^3}\right)\nonumber\\
&=&\frac{\left(\bf r_\parallel\times a_\perp+r_\perp\times a_\parallel\right)}{\gamma^3}=0.
\label{eq:rta}
\end{eqnarray}
Although $\bf r\times a$ itself is not a Lorentz invariant, we see from Eq.\ (\ref{eq:rta}) that
if there is no turning moment in one Lorentz system, there will be no turning moment in any other Lorentz system
\footnote{Although the right angular lever has usually been treated with all vectors coplanar, our analysis can be extended to the general non-coplanar case with the same conclusion\cite{tn}.}  
There is no moving lever paradox if ``tendency to rotate" is related to
$\bf r\times a$ and not ${\bf r\times}\frac{d{\bf p}}{dt}$.

We see in this example, as was emphasized in Ref.\ \cite{tn}, that caution must be observed in using pre-relativistic terms like ``force" and ``torque" in special relativity.  Although the cross product ${\bf r\times}\frac{d{\bf p}}{dt}$ is equal to the rate of change of angular momentum $\bf r\times p$, it is $\bf r\times a$ that determines whether an object will have a tendency to rotate. ``Force", too, has different meanings in special relativity.  We have used three different``forces" in this paper:
\begin{enumerate}
\item The ``Lorentz Force", $\bf F_L$ equals the rate of change of momentum $\left(\frac{d{\bf p}}{dt}\right)$. 
\item The ``Minkowski Force", ${\cal F}^\mu$, is a 4-vector, and is easy to Lorentz transform.
\item The ``Newtonian Force", ${\bf F_N}=m{\bf a}=m\frac{d{\bf v}}{dt}$ tells how an object will move. 
\end{enumerate}
These three ``forces" are the same non-relativistically, but have different applications in special relativity.
Using the wrong ``force" for a given situation, can lead to confusion, as in the right angle lever paradox.
\\
\\
{\large{\bf Appendix}}\\
\\
In this appendix, we look at the explanations of the right angle lever in references [2] and [3].
Unfortunately, the internet discussions are too numerous and wrong in too many ways to address here.
We have addressed other spurious discussions of relativistic rotation in the Appendix of Ref. [5].
In his classic text[2], Tolman considers the moving right angle lever to be ``a simple example of a stressed body in uniform translatory motion which nevertheless needs a turning moment to maintain this state of motion."
He concludes ``that the angular momentum of the system is indeed actually being increased by a flow of energy into it at exactly the rate demanded by this turning moment."  This just says that he thinks there is an explanation for why the angular momentum is increasing, but it does not address the question of how you can have 
${\bf r}\times\frac{d{\bf p}}{dt}\ne 0$ without the lever rotating.

Reference [3] comes to ``the consequent realization that in relativistic analysis there exists a net internal torque 
which exactly cancels the net external torque experienced by an extended body in dynamic equilibrium.  We find that the lever has constant angular momentum...".  Note that their conclusion of constant angular momentum directly contradicts Tolman's explanation of why the angular momentum increases.  Each of the two discussions is based on the assumption of ``internal stresses": for Tolman, to increase the angular momentum and, for Nickerson and McAdory, to keep the angular momentum constant.  Nickerson and McAdory use ``a net internal torque exactly equal and opposite to the net external torque".  But this is just the freshman confusion about the cart and the horse.  Of course the internal and external forces are equal and opposite in any rotation by Newton's third law.  We repeatedly emphasize to our students that internal forces cannot affect the motion of the body.  There is no reason this is not also true in special relativity even if Nickerson and McAdory spend seven pages doing just that.


\begin{thebibliography}{9}
\bibitem{lt}Lewis G R and Tolman R C 1909 {\it Phil. Mag.} {\bf 18} 510
\bibitem{Tolman} Tolman R C 1934 {\it Relativity Thermodynamics and Cosmology},
(Oxford at the Clarendon Press)
\bibitem{nick}Nickerson J C and McAdory R T 1975 {\it Am. J. Phys.} {\bf 43} 615.
This paper gives earlier references.
\bibitem{Spinelli}Cavalleri G, Gron O, Spavieri G, and Spinelli G 1978
{\it Am. J. Phys} {\bf 46} 108.  This paper is a comment on Ref. [3].   
\bibitem{tn}Franklin J 2006 {\it Eur. J. Phys.} {\bf 27} 1251; arXiv:physics/0603110
\end{thebibliography}
\end{document}